\documentclass[a4paper]{ESASPCS13Style}
\usepackage{epsfig}

\begin{document}

\title{Rotational evolution of very low mass objects and brown dwarfs}

\author{Jochen Eisl\"offel\inst{1} \and Alexander Scholz\inst{1}}
  \institute{Th\"uringer Landessternwarte, Sternwarte
  5, D-07778 Tautenburg, Germany}

\maketitle 

\begin{abstract}
The regulation of angular momentum is one of the key processes for our 
understanding of stellar evolution. The rotational evolution of solar-mass 
stars is mainly determined by the magnetic interaction with their 
circumstellar disk and angular momentum loss through stellar winds, and 
In contrast to solar-mass stars, very low mass (VLM) objects and brown dwarfs 
are believed to be fully convective. This may lead to major differences of 
rotation and activity, since fully convective objects may not host a 
solar-type dynamo.

Here, we report on our observational efforts to understand the rotational 
evolution of VLM objects. By means of photometric monitoring, we determined 
62 rotation periods for targets in three clusters, which form an age 
sequence from 3 to 125\,Myr. We find that VLM objects rotate faster than 
their solar-mass siblings in all evolutionary stages. Their rotational 
evolution seems to be determined by hydrostatic contraction and exponential
angular momentum loss. The photometric amplitudes of the light curves are 
much lower than for solar-mass stars. This may be explained as a consequence 
of smaller spot coverage, more symmetric spot distributions, or lower contrast
between spots and their environment. Most of these results can be explained 
with a change of the magnetic field properties with decreasing mass. VLM 
objects possibly possess only small-scale, turbulent magnetic fields.


\end{abstract}

\section{Introduction}

Rotation is one of the key parameters for stellar evolution. It is the 
parameter that -- at least in some well-behaved objects -- can be measured 
to the highest accuracy. In such objects derivations of rotation periods 
with a precision of 1 : 10000 are possible. 

In solar mass stars the investigation of their rotation has allowed us new
insights into their evolution (\cite{b95}, \cite{bfa97}, \cite{st03}). 
It has become clear that angular momentum
regulation is a direct consequence of basic stellar physics: most of the
angular momentum of a fragmenting and collapsing molecular cloud is lost in
the course of the formation of protostars. The specific angular momentum of 
protostars is, however, still one or two orders of magnitude higher than 
that of young main sequence stars. On the other hand, in their T\,Tauri 
phase solar-mass stars rotate slowly although they are accreting. The magnetic
coupling between the star and its disk, and consequent angular momentum
removal in a highly collimated bipolar jet are thought to be responsible for
this rotational braking (\cite{c90}, \cite{k91}, \cite{snowrl94}). 
After the dispersal of the disk, and thus loss of the
braking mechanism, the rotation is observed to accelerate as the stars
contract towards the zero-age main sequence (ZAMS). On the main sequence (MS)
the rotation rates of solar-mass stars decrease again because of angular
momentum loss through stellar winds.

Rotation can be investigated either by measuring stellar photospheric lines
spectroscopically, or by the determination of rotation periods from
photometric time series observations. While the former suffers from projection
effects -- the unknown inclination angle of the rotation axis with respect to
the line of sight -- the latter can be determined with high precision and
independent of inclination angle.

While a large amount of rotation periods are available in the literature for
low-mass stars in clusters younger than about 3\,Myr (ONC: \cite{herbst01}, 
\cite{hbm02}; NGC2264: \cite{l03}, \cite{lbm04}), not much is known
about the further evolution of very low mass (VLM) objects and brown dwarfs up
to the age of a few Gyr, when they are found as field objects in the solar
neighbourhood (\cite{ctc02}). Therefore, we have initiated a programme to 
obtain the required
observations, and to compare them to solar-mass stars and to evolutionary
models. For the monitoring programme we decided to follow the photometric time
series approach to obtain precise rotation periods.

In this article we will first discuss in Section 2 our targets, observations,
and data analysis. In Sect.\,3 we present our results on rotation and
variability of the sources. The observed rotation rates are then in Sect.\,4 
compared to various models of rotational evolution. Finally, Sect.\,5 contains
our conclusions.

\section{Observations and data analysis}

\begin{figure}[th]
  \begin{center}
    \resizebox{\hsize}{!}{\includegraphics[width=7cm, angle=-90]{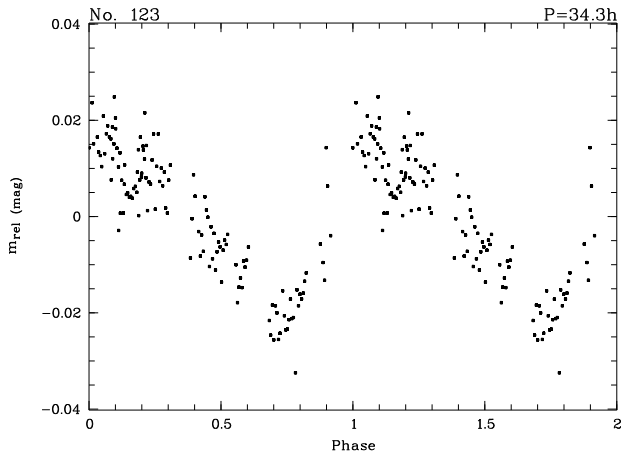}}
  \end{center}
\caption{Phase-folded light curve for a 50 M$_{jup}$ brown dwarf in the
  epsilon Ori field. The measured rotation period is 34.3\,h, and the light
  curve comprises of 129 measured data points. \label{fig1}}
\end{figure}

Since virtually no rotation periods for VLM objects and brown dwarfs with ages
older than 3\,Myr were known, it was necessary to create a period database
that complements the known rotation periods of solar-mass stars. 

In the course of our ongoing monitoring programme we have so far obtained
rotation periods for 23 VLM objects in the cluster around \object{sigma Ori} 
(\cite{se04a}), for 30 in the field around \object{epsilon Ori} 
(\cite{se04c}), which are belonging to the Ori\,OBIb association, and for 
9 objects in the \object{Pleiades} open cluster (\cite{se04b}). With ages of
about 3, 5, and 125\,Myr these three groups of VLM objects form an age
sequence that already allows us insights into a relevant part of their young
evolution. 

Our time series photometry was done with the Wide Field Imager 
(WFI) at the ESO/MPG 2.2-m telescope on La Silla in \object{epsilon Ori},
and with the CCD camera at the 1.23-m telescope at the German-Spanish
Astronomy Centre on Calar Alto (CA) in the \object{Pleiades}. The 
\object{sigma Ori} cluster was observed in two campaigns with the CCD 
cameras at the 2-m Schmidt telescope in Tautenburg (TLS) and at the 1.23-m 
telescope on Calar Alto. All time series are done in the I-band to maximise 
the observed flux from the objects. A single field was observed for sigma 
and \object{epsilon Ori}, while two adjacent fields were observed in the 
\object{Pleiades}. In sigma and \object{epsilon Ori}, we searched the 
fields prior to our monitoring campaign to identify the VLM objects from 
(I, R-I) colour magnitude diagrams, which were then further tested for 
membership using the J, H, K-band photometry from 2MASS. For the 
\object{Pleiades} we placed our fields so as to maximise the number of VLM 
objects known from the literature (e.g., \cite{pinfield00}) in them. Our 
time series photometry covers 10 nights for \object{sigma Ori}, four 
nights for \object{epsilon Ori}, and 18 nights for the \object{Pleiades}, 
in which -- weather permitting -- the respective fields were continuously 
monitored. The data were reduced following
standard recipes. Then PSF photometry of all objects in the fields was done
for the TLS and WFI data, while differential image photometry was done for the
CA data (\cite{al98}, \cite{rfg01}, \cite{gr02}). In case of PSF photometry, 
non-variable field stars are used to correct for changing airmass etc., and 
to establish a relative reference frame. Our time series analysis of this 
photometry then uses a Scargle periodogram (\cite{s82}) to find periodicity. 
The significance of this periodicity is further tested in a variety of ways 
that are described in \cite*{se04a}. Fig.\,1 shows an example for a final 
phase-folded light curve of a brown dwarf in the \object{epsilon Ori} field.

\section{Rotation and variability of VLM objects}

The general interpretation for the observed periodic variability in the light
curves of our VLM targets are surface features, which are asymmetrically
distributed on the surface and are co-rotating with the objects. Such surface
features could arise either from dust condensations in the form of ``clouds'',
or from magnetic activity in the form of cool ``spots''. Since all our objects,
because of their youth, have surface temperatures T$_{\rm eff}$ $>$ 2700\,K 
(\cite{baraffe98}) corresponding to spectral types earlier than M8, and thus 
higher than the dust condensation limits, we are most likely observing the 
effects of cool, magnetically induced spots.

\begin{figure}[th]
  \begin{center}
    \resizebox{\hsize}{!}{\includegraphics[width=7cm, angle=-90]{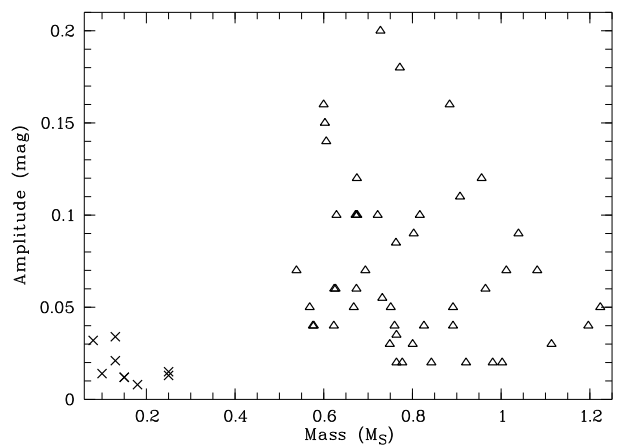}}
  \end{center}
\caption{Photometric amplitude versus mass for \object{Pleiades} stars from 
the Open Cluster Database (triangles, see Sect.\,3 for complete references) 
and our targets (crosses). The detection limit for the solar-mass stars is 
0.02\,mag, explaining the lack of stars with very low amplitudes in this 
sample. \label{fig2}}
\end{figure}

\begin{figure}[t]
  \begin{center}
    \resizebox{\hsize}{!}{\includegraphics[width=7cm, angle=-90]{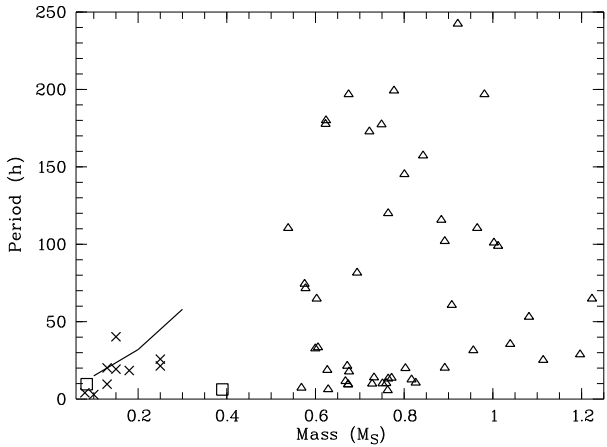}}
  \end{center}
\caption{Rotation periods versus mass in the \object{Pleiades}. Our rotation
 periods for VLM objects are shown as crosses. Triangles mark the periods for 
solar-mass stars from the Open Cluster Database. The two squares show 
periods from Terndrup et al. (1999) The solid line marks the upper 
limit to the observed $v \sin i$ values of Terndrup et al. (2000). 
\label{fig3}}
\end{figure}

It is also interesting to compare the photometric amplitudes of the periodic
variations in the light curves with those of more massive cluster
members. Such a comparison can be done well for the \object{Pleiades}, for 
which the required photometric information for solar-mass stars is available 
from the Open Cluster Database (as provided by C.F. Prosser (deceased) and 
J.R. Stauffer). 
Fig.\,2 shows than larger amplitude variations are only observed in the higher 
mass objects. It is statistically significant that the amplitude distributions 
for higher and lower mass objects are different. That only amplitudes smaller 
than 0.04\,mag are observed in the VLM objects may be attributed to the fact 
that a) the relative spot covered areas of their surfaces are smaller, b) 
their spot distributions are more symmetric or c) the spots have a lower 
relative temperature contrast with the average photosphere. 

We note that a few of the VLM objects in the two Orion regions also show large
amplitudes of up to 0.6\,mag. These variations are, however, of a more
irregular character and most likely result from hot spots originating from
accretion of circumstellar disk matter onto the object surface (see also 
\cite{fe96}).

Investigating the mass dependence of the rotation periods for the VLM and
solar-mass objects in the \object{Pleiades}, we find that their period 
distributions are also different. Fig.\,3 shows that among the VLM objects we 
are lacking members with rotation periods of more than about two days, while 
the solar-mass objects show periods of up to ten days. Although our photometric
monitoring covered a time span of 18 days, we might have missed slow rotators
among the VLM objects, if their spot patterns evolved on a much shorter time
scale, or if they did not show any significant spots. In order to investigate
these possibilities, we converted the spectroscopically derived lower limits
for rotational velocities from \cite*{terndrup99} and references therein
into upper limits for the rotation periods of the VLM objects using the radii
from the models by \cite*{chabrier97}. These rotational velocities
should not be affected by the evolution of spot patterns on the objects. The
derived upper period limits are shown in Fig.\,3 as a solid line. With a
single exception, all our data points fall below this line, and are thus in
good agreement with the spectroscopic rotation velocities. Both complementary
data sets indicate the absence of slow rotators among the VLM objects. In
fact, our data show a trend towards faster rotation even in the VLM regime
going to lower masses. A similar trend is also seen in our \object{epsilon Ori}
sample, as well as in the Orion Nebula Cluster data by \cite*{herbst01}.

\section{Rotational evolution of VLM objects}

We can now combine the periods for all three clusters, \object{sigma Ori} 
(\cite{se04a}), \object{epsilon Ori} (\cite{se04c}), and the 
\object{Pleiades} (\cite{se04b}) to try to reproduce their period 
distributions with simple models. These models should include essential
physics of star formation and evolution as described in Sect.\,1. Given the
currently available amount of information, we project the period distribution
for \object{sigma Ori} forward in time and compare consistency of the model
 predictions with our observations for \object{epsilon Ori} and the
 \object{Pleiades}. 

\begin{figure}[ht]
  \begin{center}
    \resizebox{\hsize}{!}{\includegraphics[width=7cm, angle=-90]{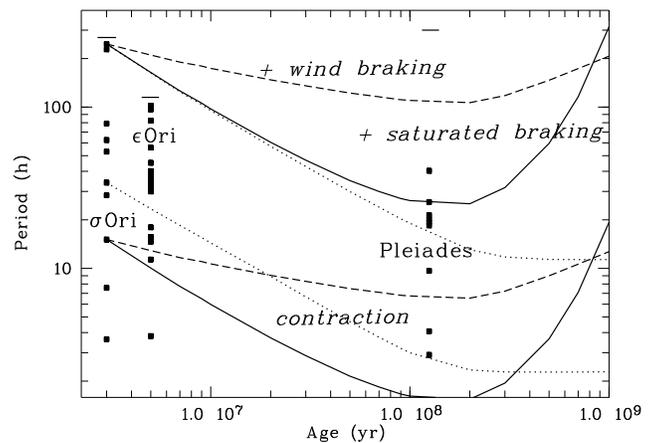}}
  \end{center}
\caption{Rotational evolution of VLM objects. The evolution of the rotation 
periods for a couple of objects for a model with hydrostatic contraction only
is shown as dotted lines. The model with additional Skumanich type wind
braking is shown as dashed lines, while saturated wind braking models are
shown as solid lines. \label{fig4}}
\end{figure}

As a first step, we take into account only the hydrostatic contraction of the
newly formed VLM objects. Changes in their internal structure may be
negligible for these fully convective objects (\cite{spt00}). In this case 
the rotation periods evolve from the initial rotation period at the age of 
\object{sigma Ori} strictly following the evolution of the radii. These radii 
were taken from the models by \cite*{chabrier97}. The dotted lines in Fig.\,4 
show the acceleration of the rotation which is coming to a halt only for ages 
older than the \object{Pleiades}, when the objects have settled. It is evident 
that this model is in conflict with the observed \object{Pleiades} rotation 
periods. Half of the \object{sigma Ori} objects would get accelerated to 
rotation periods below the fastest ones found in the \object{Pleiades} of 
about 3\,h. At the same time, even the slowest rotators in \object{sigma Ori} 
would get spun up to velocities much faster than the slower rotators in the 
\object{Pleiades}. Since the \object{sigma Ori} VLM objects surely will
undergo a significant contraction process, it is evident that significant
rotational braking must be at work until they reach the age of the 
\object{Pleiades}. 

Therefore, in a second model we add a Skumanich type braking through stellar
winds (\cite{s72}). This wind braking acts to increase the rotation periods 
$\sim$ $t^{1/2}$, see the dashed lines in Fig.\,4. According to this model, 
some of the \object{sigma Ori} slow rotators now get braked so strongly that 
they would become clearly slower rotators than are observed in the 
\object{Pleiades} (see also Sect.\,3). This indicates that even the slowest 
\object{sigma Ori} rotators seem to rotate so fast, that they are beyond the 
saturation limit of stellar winds (\cite{cdp95}, \cite{tsp00}, \cite{b03}). 
In this saturated regime, angular momentum loss is assumed to depend only
linearly on angular rotational velocity, thus rotation periods increase
exponentially with time. The solid lines in Fig.\,4 follow our model which
includes contraction and saturated wind braking. The period evolution of this
model clearly is the most consistent with our data.

For a few of our objects in \object{sigma Ori} we found evidence that they 
may possess an accretion disk. Therefore, it is interesting to explore, if 
disk-locking at young age may play a role for the evolution of rotation 
periods. Assuming disk-locking for an age up to 5\,Myr, typical for the 
occurrence of accretion disks in solar-mass stars, rotation periods would 
remain constant from the age of \object{sigma Ori} (3\,Myr) to the age of 
5\,Myr. This disk-locking scenario was
combined with the saturated wind braking, with an adapted spin-down time
scale. It is shown in Fig.\,5 as dashed lines for two objects, together with
the pure saturated wind braking model discussed above (solid lines, as in
Fig.\,4). The period evolution for both models is nearly
indistinguishable. Thus from our currently available rotation periods for
these three clusters alone, there is no strong evidence for disk-locking on
VLM objects. 
 
\begin{figure}[th]
  \begin{center}
    \resizebox{\hsize}{!}{\includegraphics[width=7cm, angle=-90]{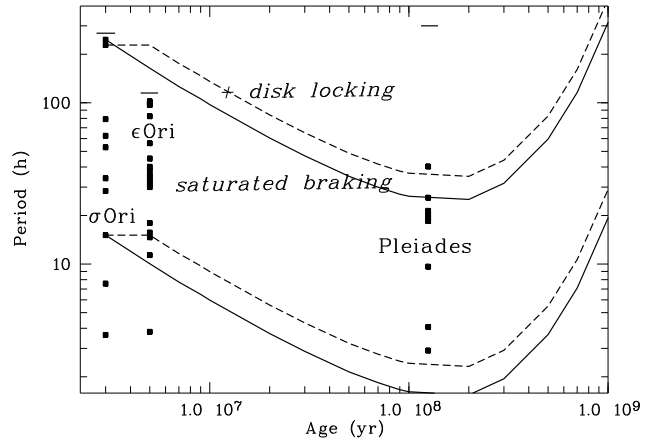}}
  \end{center}
\caption{Rotational evolution of VLM objects. The evolution of the rotation 
periods for a couple of objects for a model with hydrostatic contraction and 
saturated wind braking are shown as solid lines, as in Fig.\,4, while a model
with added disk-locking up to an age of 5\,Myr is shown as dashed lines.
\label{fig5}}
\end{figure}

\section{Conclusions}

We report results from our ongoing photometric monitoring of VLM objects -- 
so far in the clusters around \object{sigma Ori}, \object{epsilon Ori}, and 
the \object{Pleiades}, and first attempts to model their rotational evolution. 

The observed periodic variability of many VLM objects is likely caused by
magnetically induced cool spots on the surfaces of the objects. In particular
in the \object{Pleiades}, we show that variation amplitudes are lower in VLM 
objects than in solar-mass stars, indicating either less asymmetric spot 
distribution,
smaller relative spotted area, or lower contrast between spots and average
photosphere. VLM objects show shorter rotation periods with decreasing mass,
which is observed already at the youngest ages, and hence must have its origin
in the earliest phases of their evolution. 

Combining the rotation periods for all our objects, we find that their
evolution does not follow hydrostatic contraction alone, but some kind of
braking mechanism, e.g. wind braking similar to the one observed in solar-mass
stars, is required as well. Such a wind braking is intimately connected to
stellar activity and magnetic dynamo action (\cite{schatzman62}). On the 
other hand, all the 
investigated VLM objects are thought to be fully convective , and therefore
may not be able to sustain a solar-type large-scale dynamo, which is at the
heart of the Skumanich type angular momentum loss of solar-mass stars. In
fact, our modeling shows that such a Skumanich type wind braking cannot
explain our data, while saturated angular momentum loss following an
exponential braking law can. This, and the observed small photometric
amplitudes may advocate a small-scale magnetic field configuration, and may
support turbulent dynamo scenarios. Consistent theoretical models of such
dynamos that would permit rigorous testing against the observations are,
however, not yet available.

\begin{acknowledgements}

This work was partially funded by \em{Deutsche Forschungsgemeinschaft} 
(DFG), grants Ei\,409/11-1 and Ei\,409/ 11-2. 

\end{acknowledgements}

\end{document}